\documentclass[pra,twocolumn,superscriptaddress,amsmath,amssymb,aps,10pt]{revtex4-1}

\usepackage{graphicx}
\usepackage{dcolumn}
\usepackage{bm}
\usepackage{braket}
\usepackage{hyperref}
\hypersetup{
    colorlinks=true,        
    linkcolor=blue,          
    citecolor=blue,        
    filecolor=magenta,      
    urlcolor=blue           
}

\usepackage{color} 				
\definecolor{myblue}{rgb}{0.4, 0.3, 0.7}
\definecolor{dmag}{rgb}{0.6,0.0,0.6}
\definecolor{dred}{rgb}{0.8,0,0}
\definecolor{gray}{rgb}{0.5,0.5,0.5}
\usepackage[normalem]{ulem} 	

\begin{document}

\title{SU(4)-symmetric Hubbard model at quarter filling: insights from the dynamical mean-field approach}

\author{Vladyslav Unukovych}
\affiliation{Karazin Kharkiv National University, Svobody Sq. 4, 61022 Kharkiv, Ukraine}

\author{Andrii Sotnikov}
\email{a\_sotnikov@kipt.kharkov.ua}
\affiliation{Karazin Kharkiv National University, Svobody Sq. 4, 61022 Kharkiv, Ukraine}
\affiliation{Akhiezer Institute for Theoretical Physics, NSC KIPT, Akademichna Str.~1, 61108 Kharkiv, Ukraine}

\date{\today}

\begin{abstract}
We apply the dynamical mean-field approach to the four-component SU(4)-symmetric Fermi-Hubbard model to study transitions between different magnetically ordered phases as well as the hysteresis behavior in the unordered regime.
At quarter filling (one particle per site) on the square lattice we identify both the two-sublattice and plaquette-ordered antiferromagnetic phases with the corresponding entropy-driven hierarchy for critical temperatures. 
We also analyze the behavior of thermodynamic characteristics: the local double occupancy, compressibility, and entropy per particle, which are relevant for experiments with ultracold alkaline-earth(-like) atoms in optical lattices.
\end{abstract}

\maketitle

\section{Introduction}
Ultracold atomic gases in spatially periodic potentials serve as a good platform for detailed exploration of conventional systems, such as the multiorbital Hubbard model, as well as exotic states of matter with high symmetries.
In particular, with ultracold $^{173}$Yb and $^{87}$Sr atoms one can realize high-symmetry groups, such as SU($N$) with the number $N$ as high as $N=6$ and $N=10$, respectively~\cite{Wu2010, Gorshkov2010NP, Caz2014RPP}. 
The central object of the current study --- the four-component quantum system with SU(4) symmetry --- can be acquired with these atomic species by removing an excessive number of components with certain nuclear spin projections.
The remaining four internal states of an atom can be viewed then as distinct flavors or {\it pseudospin components}. 

Up to now, theoretical studies of the SU(4)-symmetric Hubbard model with respect to optical-lattice realizations have mostly focused on the case of the half-filled lowest band (see, e.g., Refs.~\cite{Wu2006MPL, Cai2013PRB, Blumer2013PRB, Yanatori2016PRB, Golubeva2017PRB}).
In contrast, due to the complexity of the respective theoretical analysis (in particular, due to the sign problem in the corresponding quantum Monte Carlo simulations), the case of average filling by one particle per site is much less explored~\cite{Szi2011EPL,Chen2016,Lee2018, Hazzard2021}, despite its significance.
For instance, in the limit of large local interaction and one particle per site the SU(4)-symmetric Hubbard model can be transformed into the corresponding Heisenberg model, which has been investigated using a number of advanced theoretical approaches (see, e.g., Refs.~\cite{FAs2005PRB, Cor2011PRL, Romen2020}).
According to some of them, in the zero-temperature limit a specific N\'eel-type arrangement of atoms can develop on the square lattice, which we call {\it plaquette order}.
At the same time, it is argued that by accounting for thermal excitations, i.e., with the entropy increase, the system can undergo a transition to the antiferromagnetic state with a conventional bipartite structure~\cite{Romen2020}.

From the experimental perspective, the low-temperature regimes in gases of alkaline-earth(-like) atoms (AEAs), e.g., $^{87}$Sr and $^{173}$Yb, are believed to become more accessible due to recent advances in cooling techniques \cite{Hofrichter16PRX, Chiu2018} and quantum gas microscopy \cite{Yamamoto16, Miranda17}.
This raises important questions about whether the many-body phases peculiar to the SU(4)-symmetric Heisenberg model remain stable with the inclusion of hopping processes and thermal excitations.
Moreover, for practical purposes not only qualitative, but also quantitative estimates and the choice of optimal physical parameters are needed to approach particular many-body regimes.

In this paper, we study how the correlations between pseudospin components, which are peculiar to the Heisenberg model, evolve by transforming to the more generic Hubbard model with SU(4) symmetry. 
We determine the stability of certain many-body regimes with respect to the change in the local interaction strength and temperature. 
Furthermore, we analyze important physical observables, which can serve as indicators of the onset of magnetic correlations and many-body phenomena in SU(4)-symmetric four-component Fermi gases in optical lattices.

\section{Model and Method}\label{sec.2}
We describe the four-component interacting Fermi gas in the framework of the Hubbard model with four internal pseudospin states and the lowest-band approximation, which corresponds to the situation in which the effective band gap is much larger than the relevant energy scales (e.g., the noninteracting bandwidth and the interaction amplitude). 
This is realized by a sufficiently strong lattice potential in all spatial directions, $V^{(\alpha)}_{\rm lat}> 5E_r$, where $\alpha=\{x,y,z\}$ and $E_r=\hbar^2 k^2/2m$ is the recoil energy of an atom with mass $m$ in the optical lattice characterized by wave number~$k$.
Below, we focus on cases, where the motional degrees of freedom in one spatial direction are suppressed while in the other two they are equal; that is, the atoms are loaded into the quasi-two-dimensional square lattice potential. Furthermore, we make the assumption that the on-site interaction $U$ can be set equal for all four components, which is valid for the ground-state manifold of AEAs, thus yielding the Hamiltonian
\begin{align} \label{eq:HubbardH}
\hat{\cal H} = &-t \sum \limits_{\langle i,j \rangle} \sum \limits_{\alpha = 1}^{4} (\hat{c}_{i \alpha}^{\dagger} \hat{c}_{j \alpha} + {\rm H.c.}) 
    - \mu \sum \limits_{j} \sum \limits_{\alpha=1}^{4} \hat{n}_{j \alpha} \nonumber\\
    &+ U\sum \limits_{j} \sum \limits_{\alpha = 1}^{4} \sum \limits_{\alpha' > \alpha}^{4}  \hat{n}_{j \alpha} \hat{n}_{j \alpha'},
\end{align}
where $\hat{c}_{i\alpha}^{\dagger}$ ($\hat{c}_{i\alpha}$) is the fermionic creation (annihilation) operator for a particle with flavor~$\alpha$ located on lattice site~$i$ and $\hat{n}_{i\alpha}=\hat{c}_{i\alpha}^{\dagger} \hat{c}_{i\alpha}$ is the corresponding number operator.
The hopping amplitude $t$ and chemical potential~$\mu$ are equal for all flavors and lattice sites.
It is worth noting that in the given form the Hamiltonian~\eqref{eq:HubbardH} is SU(4) symmetric.

The symmetry of the model can be better understood by employing the Schrieffer-Wolff transformation in the limit $t\ll U$ close to 1/4 band filling, $\sum_\alpha n_{i\alpha}\approx1$, where one can map the Hamiltonian~\eqref{eq:HubbardH} to an effective one with the same symmetry, i.e., the SU(4)-symmetric Heisenberg model,
\begin{eqnarray}
  \mathcal{\hat{H}}_{\textrm{eff}} = J
  \sum\limits_{\langle ij\rangle}
  \sum\limits_{k=1}^{15} \hat{S}_{ki}\hat{S}_{kj},
\end{eqnarray}
with positive (antiferromagnetic) exchange coupling $J=4t^2/U$.
Here $\hat{S}_{k}$ is the generator of the SU(4) symmetry group, i.e., the pseudospin projection operator to the $k$-th axis in the effective 15-dimensional spin space. 
It can be expressed through the generalized $4\times4$ Gell-Mann matrices $\boldsymbol{\lambda}_k =\{\boldsymbol{\lambda}_1,\ldots,\boldsymbol{\lambda}_{15} \}$ obeying standard commutation relations \cite{Georgi1999} similarly to the spin-1/2 case, $\hat{S}_{k} = \frac{1}{2} \hat{c}^\dag_{\alpha} \lambda_{k\alpha\beta} \hat{c}_{\beta}$. 
Therefore, the introduced pseudospin operators $\hat{S}_{k}$ can be viewed as the generators of the SU(4) group, since the operator
\begin{equation}
    \hat{\cal U}(\boldsymbol{\varphi}) 
    = \exp\left( i \sum_{k=1}^{15}\varphi_k \hat{S}_{k} \right)
\end{equation}
performs special ($\det {\cal U}=1$) unitary rotations in the corresponding space ($\varphi_k$ are arbitrary real numbers).
Now, it can be directly verified that the original Hamiltonian~\eqref{eq:HubbardH} is invariant under these transformations in the introduced pseudospin space, $\hat{\cal H}=\hat{\cal U}(\boldsymbol{\varphi}) \hat{\cal H} \hat{\cal U}^\dagger(\boldsymbol{\varphi})$.

To study thermodynamic properties and magnetic ordering phenomena, we employ the dynamical mean-field theory (DMFT) and its real-space generalization with a clustering procedure \cite{Georges1996RMP, Hel2008PRL, Sno2008NJP,Sotnikov2014PRA}.  
Although DMFT is as an approximate method for realistic lattice geometries, results obtained with this approach are valuable both for experiments and for more advanced methods, such as quantum Monte Carlo simulations, which are computationally rather demanding due to the generic presence of a sign problem for the Fermi-Hubbard model away from half filling.

To solve the Anderson impurity problem in DMFT, we employ the exact-diagonalization (ED) procedure~\cite{Caffarel1994PRL} since it is relatively fast and reliable in most regimes of interest for Fermi gases with four interacting flavors and the number of bath orbitals $n_s=4$ per each component~\cite{Golubeva2017PRB, Sotnikov2018}.
As an alternative and more accurate approach, a quantum Monte Carlo impurity solver can be successfully applied to the system under study (see, e.g., Refs.~\cite{Kotliar2002,Gul2011RMP}). 
Nonetheless, the spectral functions and quasiparticle characteristics are not the main focus of this study.
Here, we are interested in determining the phase boundaries, analyzing local observables, and computationally demanding entropy analysis, in which the ED solver remains reliable \cite{Sotnikov2015PRA,Golubeva2017PRB, Sotnikov2018}.

\begin{figure}
\includegraphics[width=\linewidth]{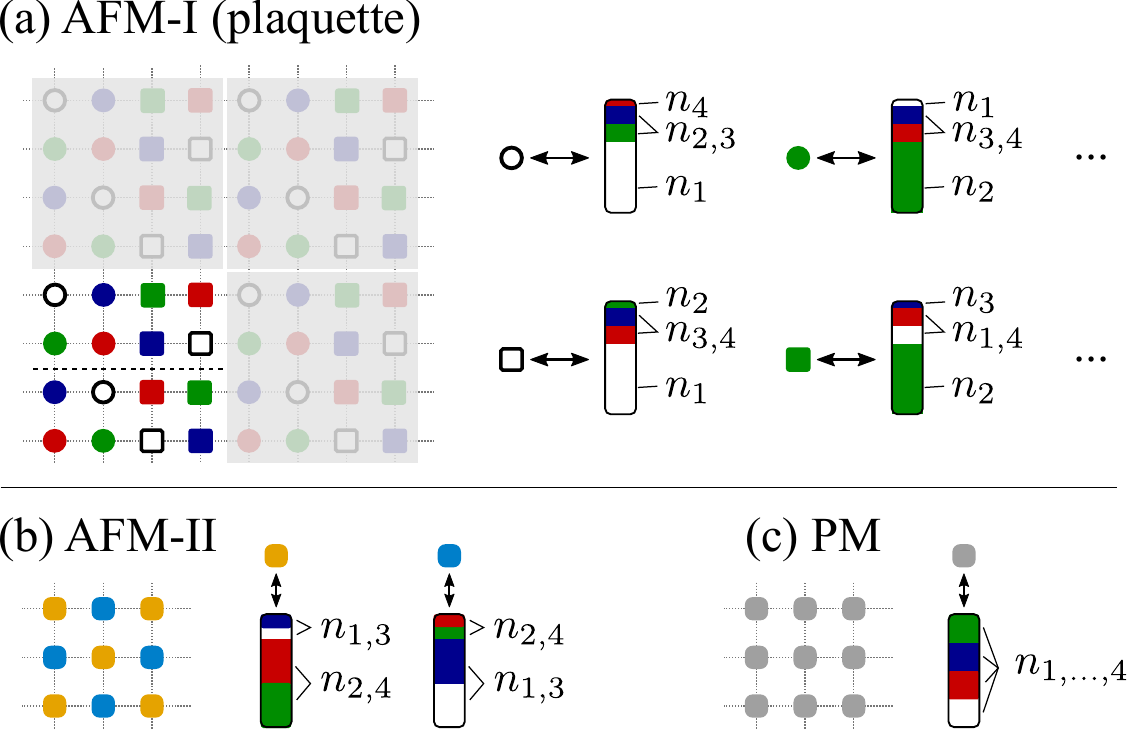}
    \caption{Spatial configurations of local occupancies on the square lattice: (a) plaquette-ordered state (the shaded regions are the same patterns that can be generated by regular lattice translations), (b) two-sublattice AFM, and (c) the unordered paramagnetic (PM) state.
    In the interpretation of the colored symbols, the length of the whole vertical bar corresponds to $n=1$, while its relative filling by different colors indicates the fraction $n_\alpha$ of the pseudospin components on the lattice site.
    The two different shapes of symbols in (a) correspond to two possible types of permutations of components with minor occupancies (see the text for details).
    }
    \label{fig:configs}
\end{figure}

To obtain the lattice Green's functions~$G_\sigma$ for self-consistency conditions in DMFT, one can express them in terms of the noninteracting density of states for a particular lattice geometry and the local self-energies~$\Sigma_\sigma$ calculated within the impurity solver. 
However, this approach works well only in the homogeneous or bipartite-ordered, i.e., two-sublattice, antiferromagnetic (AFM) regimes.
In more general situations, e.g., the plaquette-ordered AFM state [see Fig.~\ref{fig:configs}(a)], we employ the real-space generalization with clustering procedure as, e.g., described in Ref.~\cite{Sotnikov2015PRA}.
In particular, for the SU(4)-symmetric system under study we solve impurity problems on 16 lattice sites in parallel (a plaquette of the size of $4\times4$ sites). 
After that, we use regular translations along crystallographic axes for the self-energy values to the remaining sites of the system (typically, it is sufficient to restrict calculations to $64\times64$ sites to exclude any impact of the finite-size effects in the regimes of interest).
The choice of the size and shape of the plaquettes is suggested by DMFT convergence in the large system ($64\times64$ sites) from random initial conditions for the Anderson parameters on all lattice sites. 
In principle, as soon as a specific structure of the ordered configuration becomes established, the number of independent calculations of the impurity problem can be further reduced by using additional symmetry considerations.

Within the introduced real-space extension of the DMFT approach, the corresponding lattice Green's function is obtained from the inversion of the following real-space matrix:
\begin{eqnarray} \label{eq.Nsub}
    [{\bf G}^{-1}_{\sigma}(i\omega_n)]_{ ss'} 
    = [i \omega_{n} + \mu 
    - \Sigma_{\sigma s}(i \omega_n)]\delta_{ ss'} - t_{ ss'},
\end{eqnarray}
where the indices $s$ and $s'$ denote the lattice sites, $\delta_{ ss'}$ is the Kronecker symbol, $\omega_n = \pi(2n+1)/\beta$ is the fermionic Matsubara frequency, and $\beta$ is the inverse temperature ($\beta=1/k_BT$).
The hopping matrix element $t_{ ss'}$ equals $t$ if tunneling is possible between lattice sites $s$ and $s'$ and zero otherwise, which depends on the lattice geometry, system size, and boundary conditions of the original model~(\ref{eq:HubbardH}).

As soon as the DMFT convergence criteria are fulfilled, the set of local observables is evaluated on each lattice site. 
Unless specified otherwise, below we employ the units of the hopping amplitude~$t$ in all energy-related quantities.

\section{Results}\label{sec.3}
\subsection{Order parameters and phase diagram}
In the numerical analysis of the Hubbard model at quarter filling we observe two types of AFM-ordered states: AFM-I and AFM-II, which are schematically illustrated in Figs.~\ref{fig:configs}(a) and \ref{fig:configs}(b), respectively.
The low-temperature AFM-I configuration can be viewed as a generalization of the N\'eel-type dimer order in the SU(4)-symmetric Heisenberg model~\cite{Cor2011PRL}.
Indeed, the local configurations, differing only by their minority contributions [denoted by the same color but different symbol shapes of symbols in Fig.~\ref{fig:configs}(a)], become equal in the limit $U\to\infty$.
But as soon as the fluctuations originating from the hopping processes become sizable, the minority contributions form an additional modulation.
Let us clarify the structure and origin of this observation below.

In the AFM-I phase, to  determine the flavor which has the least average filling on the given site for a system with a square lattice geometry, it is sufficient to identify the flavor that has dominant contributions in two out of four nearest-neighbor sites.
In particular, let us choose the lattice sites which are dominantly occupied by the first component, $n_1\sim1$, denoted by white symbols in Fig.~\ref{fig:configs}(a). 
These sites have a pair of nearest neighbors with  either $n_4\sim1$ (red symbols) or $n_2\sim1$ (green symbols).
Due to flavor-selective Pauli blocking of hopping processes, the local occupancy of the corresponding minority component ($n_4$ or $n_2$, respectively, on the sites with $n_1\sim1$) becomes suppressed more than other occupancies. 
This results in two types of hierarchies of local fillings denoted by different brackets,
    \[ 
    (n_1):~n_1>n_{2,3}>n_4,\quad
    [n_1]:~n_1>n_{3,4}>n_2.
    \]
which are encoded by circles and squares in Fig.~\ref{fig:configs}(a), respectively.
While there are three combinations of minority occupancies possible (e.g., $\langle n_1\rangle :~n_1>n_{2,4}>n_3$ is the third one), according to our observations, it is sufficient to have two types of the described modulations to cover the whole plaquette in a closed and consistent way.

In the first place, we have to determine the boundaries between the above-mentioned AFM and paramagnetic (PM) phases. 
For this purpose we introduce two positive-valued order parameters, the ``magnetizations'' $m_1$ and $m_2$, as linear combinations of local occupancies~$n_i$,
\begin{eqnarray}\label{eq:ordpar}
      &&m_1=n_{1*}-n_{2*}, \nonumber
      \\
      &&m_2=n_{1*}+n_{2*}+n_{3*}-3n_{4*},
\end{eqnarray}
where the asterisk means that the flavors are permuted to yield the descending order of the respective values, i.e., $n_{1*}\geq\ldots\geq n_{4*}$.
Note that this is one of many possibilities to ascribe two independent order parameters to the phases under study. Obviously, with the two conditions for the local occupancy, $\sum_{\alpha=1}^{4}n_{\alpha}=1$ and $n_\beta=n_\gamma$ (the flavors $\beta$ and $\gamma$ are different; they must form, at the least, a pair, as in the AFM-I phase), the number of independent variables decreases from four to two.

\begin{figure}
\includegraphics[width=\linewidth]{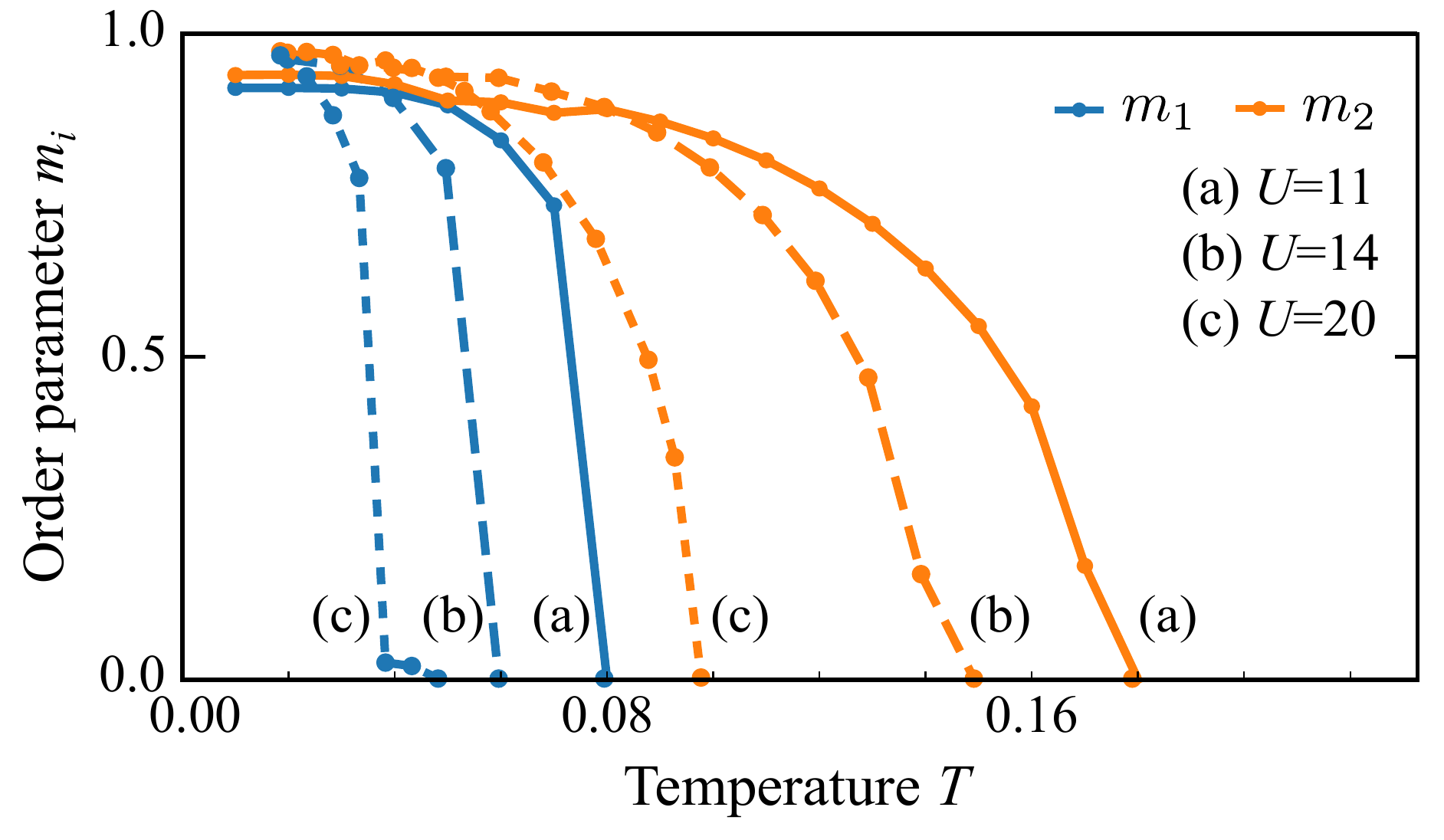}
    \caption{Temperature dependencies of the local order parameters introduced in Eq.~\eqref{eq:ordpar} at different interaction strengths~$U$: (a) $U=11$, (b) $U=14$, and (c) $U=20$.}
    \label{fig:ordpar}
\end{figure}

At variable local interaction strength~$U$ and temperature $T$, we obtain dependencies similar to those shown in Fig.~\ref{fig:ordpar}.
The magnetization behavior $m_2(T)$ indicates that the transition from the AFM-II to PM phase is continuous, i.e., of the second order.
At the same time, the transition from the plaquette-ordered AFM-I to AFM-II phase is of the first order.
We come to the latter conclusion not only from the dependence of the magnetization~$m_1(T)$ itself, which should be viewed as supplementary in this context, but mainly from the symmetry arguments.
While comparing the structures of the AFM-I and AFM-II phases in Fig.~\ref{fig:configs}, we can note that no flavor (i.e., no group of symbols with the same color) in the AFM-I phase can be viewed as a host for the checkerboard-type sublattice arrangement required for the AFM-II modulation.
In other words, while in half of the lattice sites the dominant component can be admixed with another flavor in a continuous way to approach the AFM-II regime, in the other half the relative occupancies must be simultaneously exchanged between flavors.
Note that similar symmetry considerations and hierarchies of phase transitions of different types also appear in the SU(3)-symmetric Hubbard model at 1/3 band filling \cite{Sotnikov2015PRA}.

\begin{figure}
\includegraphics[width=\linewidth]{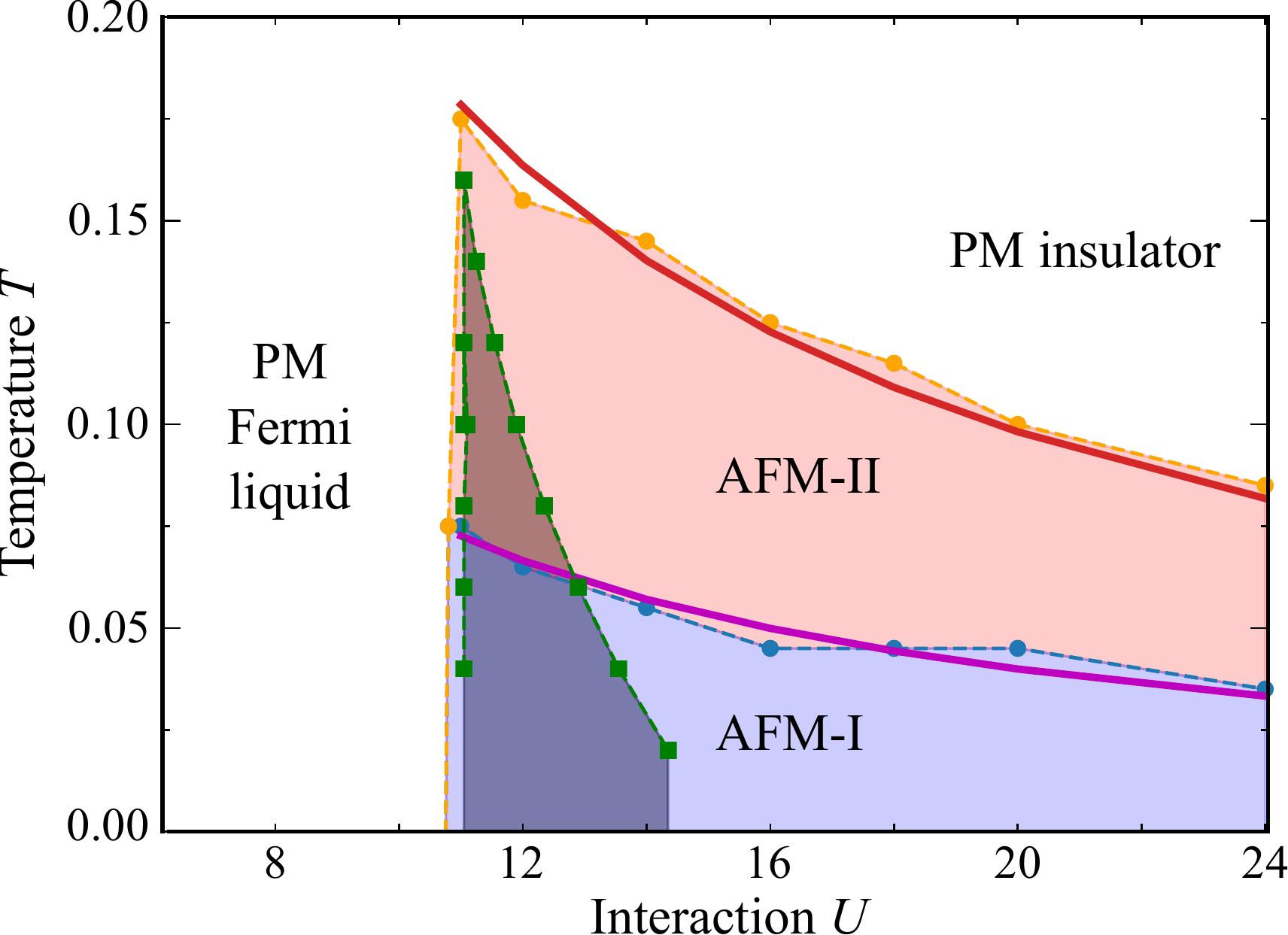}
    \caption{Phase diagram of the SU(4)-symmetric Hubbard model at quarter filling.
    The symbols connected by thin lines correspond to DMFT data, while the thick solid lines are the least-squares fits to the $T_c\propto t^2/U$ dependence, with the coefficients $0.79$ and $1.96$ for the lower and upper phase boundaries, respectively. The dark shaded area is the metal-insulator hysteresis region obtained under the PM constraint.
    }
    \label{fig:pd}
\end{figure}

Next, we construct the phase diagram shown in Fig.~\ref{fig:pd}.
It clearly points to the absence of the ordering mechanism at weak coupling $U/t\lesssim10$, i.e., in the Fermi liquid regime, where the system remains SU(4) symmetric down to zero temperature.
However, as soon as the system undergoes the metal-insulator transition (the shaded region in Fig.~\ref{fig:pd}), the AFM correlations between localized moments become crucial.
We note that there is an entropy-driven hierarchy of critical temperatures.
Obviously, the AFM-II phase has a higher entropy capacity but cannot remain stable down to zero temperature due to the third law of thermodynamics.
In contrast, if the whole lattice can be covered by finite-size plaquettes as in the AFM-I phase [see Fig.~\ref{fig:configs}(a)], this many-body configuration becomes consistent with the third law of thermodynamics.
In this respect, one could also draw an analogy to the SU(3)-symmetric Hubbard model at $n=1$ \cite{Sotnikov2014PRA,Sotnikov2015PRA}.

We should also note that according to our observations the AFM-I to AFM-II transition, in principle, can be accompanied by hysteresis features. However, in the DMFT analysis, we usually perform calculations starting from the phase with lower symmetry (AFM-I) and gradually increase the temperature.
Within this approach, at a certain temperature $T_{c1}$ the DMFT algorithm ceases to converge to the AFM-I configuration but naturally finds the AFM-II solution with stable convergence on all lattice sites.
Therefore, the AFM-I to AFM-II transition line in Fig.~\ref{fig:pd} corresponds to the upper boundary~$T_{c1}^{>}$. In contrast, the determination of the lower boundary $T_{c1}^{<}$ depends on specific (symmetry-breaking) adjustments of the DMFT iterative procedure itself, meaning that it cannot be accurately determined within this study.

\subsection{Metal-insulator transition and relevant observables}
\begin{figure}
\includegraphics[width=\linewidth]{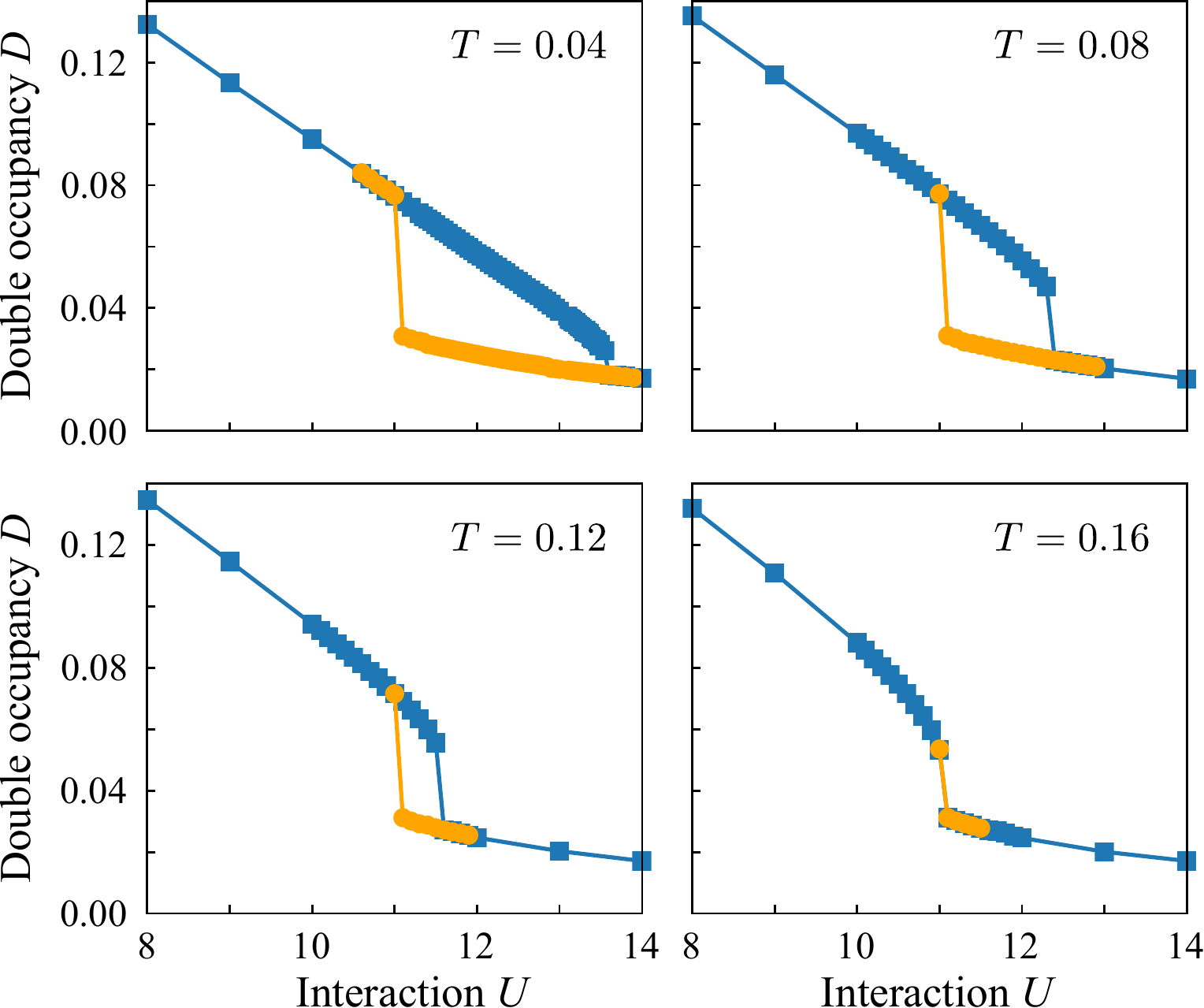}
    \caption{Hysteresis behavior of the double occupancy as a function of the interaction strength at constant temperature under the PM constraint in DMFT.
    }
    \label{fig:Doub}
\end{figure}

The employed theoretical approach allows us to analyze the behavior of local correlation functions. 
In particular, the double occupancy $D_j=\sum_{\alpha'>\alpha}\langle \hat{n}_{j\alpha}\hat{n}_{j\alpha'}\rangle$ is the one that can be effectively measured in cold-atom experiments \cite{Tai2012Nat}.
Following the standard theoretical procedure, to exclude the impact of magnetic correlations, we artificially impose the translational symmetry on all lattice sites by the corresponding (paramagnetic) self-consistency conditions of DMFT. 
The observed behavior at constant temperature and a continuous increase (decrease) of the interaction strength is shown in Fig.~\ref{fig:Doub}.
The corresponding region with coexisting metallic and insulating solutions is also indicated in Fig.~\ref{fig:pd}.

Note that one can typically suppress the magnetic correlations by introducing a hopping process across one of the diagonals, i.e., by a minor change in the lattice geometry towards the triangular one.
In this way the hysteresis behavior in the paramagnetic regime can be more effectively decoupled from the onset of magnetic correlations.

\begin{figure}
\includegraphics[width=\linewidth]{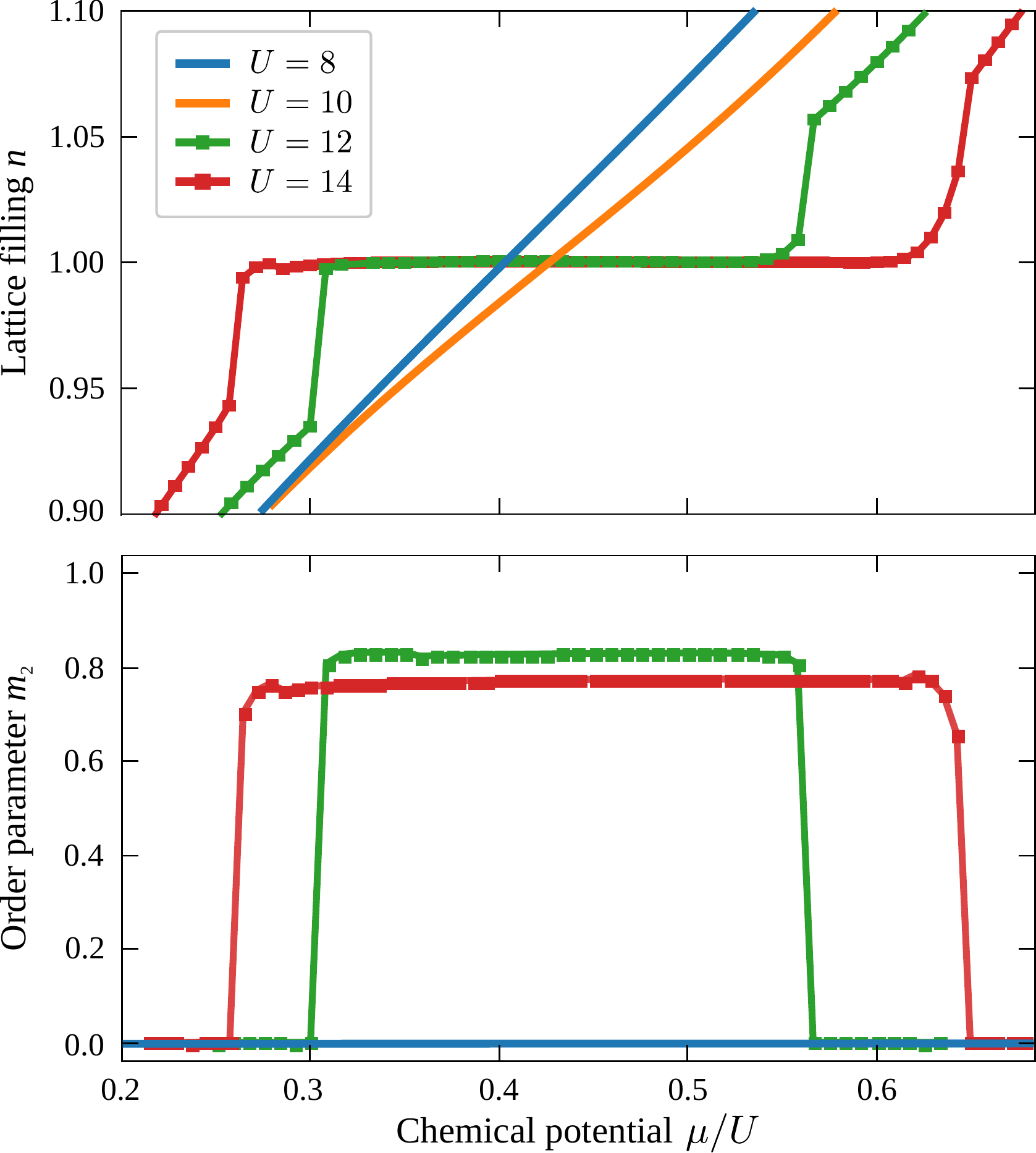}
    \caption{Lattice filling~$n$ and order parameter $m_2$ as functions of the chemical potential (in units of the respective interaction strength~$U$) at $T=0.1$ and different $U$. At $U=12$ and $U=14$ the plateau structures are accompanied by the AFM-II ordering.
    }
    \label{fig:Compr}
\end{figure}

The isothermal compressibility $\kappa_T=n^{-2} ({\partial n}/{\partial \mu})_{T}$ is another accessible observable in cold-atom experiments. 
It can be measured through the variation of the trap curvature and subsequent analysis of the changes in the density profiles (see, e.g., {Refs.~\cite{Gorelik2009,Hofrichter16PRX}}).
In Fig.~\ref{fig:Compr} we compare density distributions at four different interaction strengths and fixed temperature.
They clearly indicate the formation of plateau structures in the Mott-insulating regime.
According to Fig.~\ref{fig:Compr}, we also observe that the corresponding values of $\kappa_T$ at $T=0.1$ across the metal-insulator transition drop by around two orders of magnitude, e.g., $\kappa_T\approx6\times10^{-2}$ and $\kappa_T\approx4\times10^{-4}$ at $U=10$ and $U=12$, respectively.
By means of the local-density approximation the depicted dependencies can also be transformed to the spatial density distributions in the trap, e.g., $n(r)$ by taking $\mu(r)=\mu_0-V_{0}r^2$ in the case of the isotropic harmonic trapping potential with amplitude~$V_0$.
Note, however, that this approximation neglects the proximity effects in the trap, which result in additional softening of the  Mott-plateau features on the edges (see, e.g., Ref.~\cite{Sotnikov2013PRA}). This can be accounted for within the introduced real-space extension of DMFT but goes beyond the scope of the current study.

\subsection{Entropy analysis}
\begin{figure}
\includegraphics[width=\linewidth]{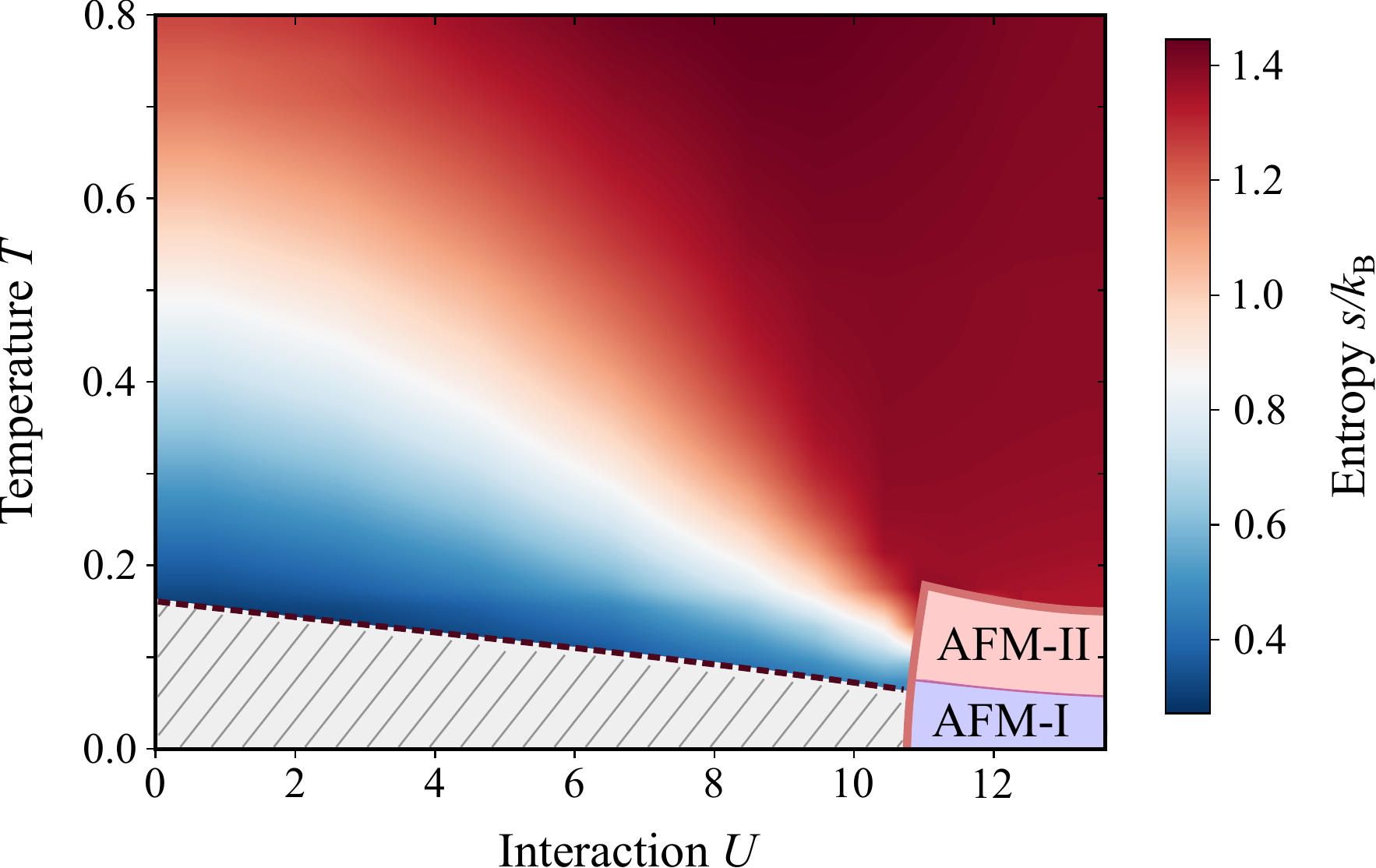}
    \caption{Contour plot of the entropy per particle as a function of the temperature and interaction strength in the SU(4)-symmetric Hubbard model at $n=1$. The ordered phases are taken from Fig.~\ref{fig:pd}, while the hatched region indicates the limitations of the numerical approach with no reliable quantitative data for the entropy.
    }
    \label{fig:Entr}
\end{figure}

Within access to the density distribution $n(\mu)$ of the interacting Fermi gas as a function of the chemical potential (see also Fig.~\ref{fig:Compr}), one can determine the entropy per particle $s$, which can be viewed as a more relevant thermodynamic quantity (compared to the temperature~$T$) in cold-atom experiments.
To this end, we employ the Maxwell relation in the integral form,
$s=\int_{-\infty}^{\mu_{1}}\left(\frac{\partial n}{\partial T}\right)d\mu$, 
where the upper limit~$\mu_1$ is determined from the condition $n(\mu_1)=1$.
Note that this approach requires evaluations of derivatives, which can be a source of numerical noise in the entropy analysis. However, in the regimes of interest these remain smooth functions.  
The low-$T$ limitations indicated in Fig.~\ref{fig:Entr} (the hatched area) correspond to the regime of the appearance of systematic errors. In this regime, the numerically obtained values for the entropy cease to be monotonously decreasing functions of the temperature. We attribute this limitation to the finite number of bath orbitals employed in the ED solver. 

According to Fig.~\ref{fig:Entr}, in the region of weak and intermediate coupling, the well-known Pomeranchuk cooling effect is observed \cite{Tai2012Nat}; that is, at fixed entropy the temperature decreases with increasing interaction strength. For the SU(4)-symmetric Fermi gas at quarter filling, the critical entropy value at which the estimated peak of the AFM-II phase can be reached is $s_c^{\rm max}\approx1.3k_{\rm B}$ (see Fig.~\ref{fig:Entr}).
This value significantly exceeds the corresponding critical value for AFM ordering in the SU(4)-symmetric Hubbard model at half filling ($n=2$) \cite{Golubeva2017PRB} and the critical value for orbital ordering in the four-component [SU(2)$\times$SU(2)]-symmetric $^{173}$Yb Fermi gas at $n=1.5$ \cite{Sotnikov2020PRR}.
Therefore, according to typical experimental limitations in the entropy per particle, the observation of magnetic correlations in the SU(4)-symmetric Hubbard model at quarter filling looks advantageous.

Note that the calculated entropy values correspond to the bulk properties of the system. In fact, we also verified that within a direct DMFT analysis, in the trapped system metallic shells with $n<1$ possess higher entropy per particle. 
This means that in the case one realizes the inhomogeneous system with $n\approx1$ in the central region to approach magnetic ordering, the estimated theoretical bounds for the total entropy per particle $S_{\rm tot}/N$ can be further increased.

\section{Conclusion}\label{sec.4}
We studied low-temperature characteristics of four-component Fermi gases in optical lattices described by the SU(4)-symmetric Hubbard model at quarter filling.
The model itself can govern not only the low-temperature behavior of cold-atom systems, but also certain classes of crystalline materials \cite{Li1998PRL,Yamada2018}.

The theoretical analysis was developed in the framework of the dynamical mean-field theory.
It was shown that in quasi-two-dimensional square lattice geometry the system undergoes the entropy-driven sequence of phase transitions as soon as the local interaction between atoms is sufficient for the onset of the insulating regime, i.e., at $U\gtrsim11t$.
The observed plaquette ordering in the system under study can be viewed as a generalization of the limiting case -- the N\'{e}el-ordered dimer configuration analyzed in the framework of the SU(4)-symmetric Heisenberg model~\cite{Cor2011PRL}.
At the same time, the studied two-sublattice AFM-II phase with higher entropy capacity emerges similarly to how it emerges in the SU(3)-symmetric Hubbard model~\cite{Sotnikov2014PRA} or in SU($N\geq3$)-symmetric Heisenberg models~\cite{Romen2020}.

By means of DMFT we also analyzed the metal-insulator hysteresis region and identified that its temperature range practically coincides with the range for magnetic ordering, in contrast to the Hubbard model with two or three interacting components. 
Our entropy analysis suggests that this first-order metal-insulator phase transition region (in contrast to the high-$T$ crossover regime) can be accessed in current and near-future experiments with alkaline-earth-like atoms in optical lattices.

In addition to measuring magnetic correlations between different fermionic flavors and analyzing related structure factors, we suggested other related thermodynamic quantities as potential indicators of phase transitions and many-body phenomena in the system under study.
In particular, the double occupancy and compressibility also show clear signals across the phase boundaries. 
The obtained entropy dependencies provide quantitative information on how the system can be effectively cooled by means of the Pomeranchuk effect and whether the phases under study can be effectively approached starting from certain  initial values of the gas temperature.

The impact of thermal fluctuations is an additional factor that should be properly accounted for. 
Given the limitations of the DMFT approach itself \cite{Georges1996RMP}, the obtained results should be viewed as approximate for quasi-two-dimensional lattice geometries and must be correctly referred to finite-range correlations, which are the main focus in cold-atom experiments.
Therefore, from the theoretical perspective, 
it is important to analyze magnetic correlations and thermodynamic quantities by means of recent developments in more advanced theoretical techniques (see, e.g., Refs.~\cite{Maier2005,Rohringer2018,Hazzard2021,DelRe2021}).

\begin{acknowledgments}
The authors acknowledge support from the National Research Foundation of Ukraine, Grant No.~0120U104963 and the Ministry of Education and Science of Ukraine, Research Grant No.~0120U102252.
Access to computing and storage facilities provided by the Poznan Supercomputing and Networking Center (EAGLE cluster) is greatly appreciated.
\end{acknowledgments}

\bibliography{SU4_quarter}	
\end{document}